\documentclass[aps,prl,twocolumn,groupedaddress,showkeys]{revtex4}
\usepackage{epsfig}

\begin{document}


\title{Macromolecular unfolding properties in presence of compatible solutes}

\author{Jens Smiatek$^1$}
\author{Hans-Joachim Galla$^2$}
\author{Andreas Heuer$^1$}
\affiliation{$^1$Institute of Physical Chemistry, University of Muenster, D-48149 M{\"u}nster, Germany\\
	     $^2$Institute of Biochemistry, University of Muenster, D-48149 M{\"u}nster, Germany}


\begin{abstract}
We present Molecular Dynamics simulations of Chymotrypsin inhibitor II and PEO in presence of compatible solutes. Our results indicate that the native compact state of the studied macromolecules is stabilized 
in presence of hydroxyectoine. We are able to explain the corresponding mechanism by a variation of the solvent properties for higher hydroxyectoine concentrations. 
Our results are validated by detailed free energy calculations.
\end{abstract}

\date{\today}

\maketitle
\section{1. Introduction}
The cell metabolism of microbacteria at extreme temperatures and dry environments is protected by the presence of compatible solutes 
which are also called osmolytes, co-solvents or extremolytes \cite{Lentzen06,Morris78,Driller08,Galinski85,Galinski93}.
It has been found that compatible solutes are strongly hygroscopic and prevent denaturation processes of the cell proteins by
increasing the melting temperature \cite{Lentzen06,Driller08,Smiatek11}. One important osmolyte which is often used in industrial products is hydroxyectoine \cite{Lapidot88,Lapidot93}.
Especially in the last years, several studies have been published to investigate the general function of osmolytes. Most of the theories which aim to describe the stabilization mechanism accurately 
can be divided into direct and indirect 
interactions \cite{Lee81,Arakawa83,Arakawa85,Timasheff01,Timasheff02,Galla10,Yu04,Yu07,Rose2006,Schellman2003,Bolen2006,Rose2008,Tolan2011,Gao10,Daggett07,Daggett04,Murphy02,Patey08,Timasheff97,Pettitt10,Grubmuller07,Berne08,Bolen07,Sironi11,Sadowski11,Bolen07a,Roesgen05,Roesgen09}. 
In the context of extremolyte protein stabilization, we will briefly discuss in the following the most common ones.\\
The framework of the preferential exclusion model as an indirect mechanism is established by the consideration of a protein dissolved in aqueous solution that contains compatible 
solutes \cite{Lee81,Arakawa83,Arakawa85,Timasheff01,Timasheff02,Galla10,Yu04,Yu07}. 
Within this model it is assumed that the chemical potential of the water at the proteins surface is perturbed by the presence of the co-solute. The corresponding thermodynamic interactions \cite{Timasheff02} lead to 
deviations from the bulk molar concentration of the co-solvent compared to the local environment around the protein.
These deviations can be related to a preferential binding or a preferential exclusion of the co-solute.\\ 
The dynamic exchange of extremolytes in the local surrounding of the proteins surface result in free energy differences due to exclusion. These differences are 
compensated by a significant increase of the 
water molecules in the local environment of the protein. Preferential hydration and preferential exclusion are therefore 
synergetic effects which appear simultaneously. According to the original idea, the stabilizing effect on the protein is mainly caused by the excess solvent molecules
whose interactions restrict the protein to its native state by increasing the melting temperature \cite{Yu04,Yu07}.
A varying melting temperature as well as a strong influence on the water structure in presence of ectoine, trimethylamine-N-oxide, amino acids and sugars has been also observed in 
experiments \cite{Lee81,Knapp99,Yancey81,Smith04,Winter11}.
In addition, the preferential exclusion model has been used to understand the effects of monolayer structure formation \cite{Galla10}.\\
Molecular Dynamics simulations have indicated \cite{Yu04,Yu07} that the preferential exclusion model is relevant for a specific class of proteins.
It has been found that the co-solvents do not directly interact with the proteins surface in agreement to the theory, but
slow down the diffusion of solvent molecules in the local environment of the protein \cite{Yu04}. It has been concluded that this restriction is responsible for the change of the melting temperature 
and therefore the sustainment
of the kinetic stability of the protein.
This has been in particular investigated for the Chymotrypsin inhibitor II (CI2) \cite{Yu07}.  
Other proteins like Met-Enkephalin show deviations to this model which result in the authors conclusion, that the preferential exclusion model is only relevant for proteins with well-defined 
and pronounced hydration 
shells \cite{Yu07}. Macromolecules which are less well-hydrated, {\em e. g.} Met-Enkephalin are only slightly influenced in their unfolding processes such that the melting temperature remains unchanged.\\
A different mechanism in contrast to the preferential exclusion assumption is proposed by the transfer free energy model 
\cite{Rose2006,Rose2008} which relates the increased stability of the protein to a solvent quality decrease in presence of compatible solutes.
This leads to favorable intra-molecular hydrogen bonds compared to unfavorable solvent-protein interactions. The excess of intramolecular hydrogen bonds results in an increased stability of the native structure
which is validated by a shifting of the equilibrium constant to the folded state.\\
In this paper we focus on the properties of CI2 and PEO (polyethylene oxide) in presence of compatible solutes.
Our results indicate that preferential hydration
is absent for these molecules although an increased stability of the native state can be observed.
We have found that the free energy landscape is significantly changed in presence of hydroxyectoine such that unfolding becomes 
energetically less favorable. In addition we present numerical findings which validate that changes of the solvent properties are mainly responsible for the observed stabilization behavior.\\ 
The paper is organized as follows. In the next sections we introduce the theoretical background and the numerical details. The results are discussed in the fourth section. 
Our conclusions and a brief summary are presented in the fifth section.
\section{2. Theoretical background}
Structural properties of proteins and polymers can be described by the end-to-end radius $r_e$ \cite{Doi86,deGennes,Flory}.
The end-to-end vector is given by
\begin{equation}
\vec{r}_e=\vec{r}_N - \vec{r}_1
\end{equation}
where $\vec{r}_{N/1}$ denotes the position of the $N$-th and the first monomer. 
The end-to-end radius obeys a scaling behavior which is expressed by 
\begin{equation}
r_e^2\sim N^{2\nu}
\label{eq:re}
\end{equation}
where $\nu$ denotes the excluded-volume parameter \cite{Doi86,deGennes,Flory} which is
dependent on the solvent properties \cite{Doi}. Hence a good solvent for the macromolecule results in a swollen conformation whereas a bad solvent leads to a shrinkage. \\
A general way to calculate free energy differences for macromolecular conformations or solvation properties in computer simulations is
thermodynamic integration \cite{Frenkel96}. 
The differences can be computed by the coupling of the Hamiltonian $\mathcal{H}$ to a perturbation parameter $\lambda$ which corresponds to a well-defined state \cite{Kollman93}.
This parameter regulates the strength of the conservative
interactions from fully developed to absence. The free energy difference between state A and state B
is defined by
\begin{eqnarray}
\Delta F_{AB} &=& F(A)-F(B) = \int_{\lambda_A}^{\lambda_B}\frac{\partial F(\lambda)}{\partial \lambda}d\lambda\nonumber\\
& & = \int_{\lambda_A}^{\lambda_B}\left<\frac{\partial\mathcal{H}(\lambda)}{\partial \lambda}\right>d\lambda
\end{eqnarray}
where $\lambda$ is in the parameter range from $0$ to $1$.
\begin{figure}[h!]
 \includegraphics[width=0.5\textwidth]{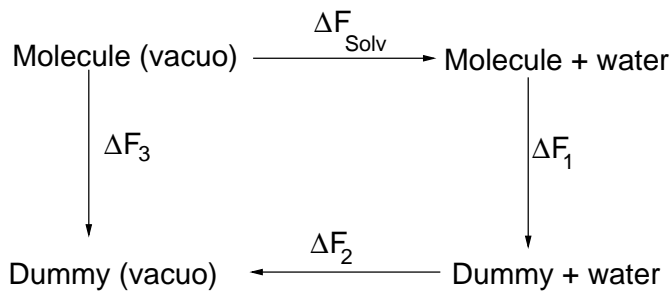}
\caption{Thermodynamic cycle for the calculation of solvation free energy differences.
}
\label{fig1}
\end{figure}
This equation can now be used to calculate free solvation energies \cite{Kollman93,Macedo10,Leach01} where $\lambda=0$ represents the dummy state without intra- and intermolecular interactions.
The corresponding thermodynamic cycle is shown in Fig.~\ref{fig1}. The
solvation free energy is then defined by
\begin{equation}
\Delta F_{_{solv}} = \Delta F_3- \Delta F_1
\end{equation}
where $\Delta F_2$ vanishes due to non-existing interactions. 
\section{3. Numerical details}
We have performed Molecular Dynamics simulations in explicit SPC/E water model solvent \cite{Straatsma87} with the software package GROMACS \cite{Berendsen95,Hess08,Spoel05}
and variations of the the force field GROMOS96 \cite{Oostenbrink04}. 
We created the chemical structure of hydroxyectoine ((4S,5S)-2-methyl-5-hydroxy-1,4,5,6-tetrahydropyrimidine-4-carboxylic acid) as it was in detail described in Ref.~\cite{Smiatek11}.\\ 
The Molecular Dynamics simulations of Chymotrypsin inhibitor II (CI2) have been carried out in a
cubic simulation box with periodic boundary conditions. The box consisted of
$(5.76637\times 5.76637\times 5.76637)$ nm$^3$ filled with 5498 SPC/E water molecules. The presence of hydroxyectoine was simulated by a 0.087 mol/L solution which corresponds to 10 molecules.
CI2 was taken from the pdb entry 1YPB \cite{Henrick94}.\\
The polyethylene oxide (PEO) chains consisted of 31 monomers where the topology of the chain was modeled by the PRODRG server \cite{vanAalten04}.
The box has a side length of $(5.32547\times 5.32547\times 5.32547)$ nm$^3$ filled with 4944 water molecules. The presence of hydroxyectoine was simulated for several concentrations 
between 0 to 0.11 mol/L which corresponds to ten molecules.\\
Electrostatic interactions have been calculated by the Particle Mesh Ewald sum \cite{Pedersen95}.
The time step was $\delta t=2$ fs and the temperature was kept constant by a Nose-Hoover thermostat \cite{Frenkel96}   
at 300 K for the PEO molecule and 400 K for the Chymotrypsin inhibitor II. 
All bonds have been constrained by the LINCS 
algorithm \cite{Fraaije97}.
After minimizing the energy, we performed 1 ns of equilibration followed by a 10 ns simulation sampling run.\\
Solvation free energy calculations for ectoine ((S)-2-methyl-1,4,5,6-tetrahydropyrimidine-4-carboxylic acid) have been performed by independent simulation runs for parameter values of
$\lambda \in$ $\{$ 0, 0.05, 0.1, 0.15, 0.2, 0.25, 0.3, 0.35, 0.4, 0.45, 0.5, 0.55, 0.6, 0.65, 0.7, 0.75, 0.8, 0.85, 0.9, 0.95, 1$\}$.
Each configuration has been equilibrated for 250 ps and the computation of free
energy values has been achieved in a 900 ps simulation run for each $\lambda$.\\
To avoid singularities, the ordinary potentials have been replaced
by the following
soft core interaction
\begin{eqnarray}
V_{sc} &=& (1-\lambda)V^A\left[(\alpha\sigma^6\lambda+r^6)^{\frac{1}{6}}\right]\nonumber\\
& & +\lambda V^B\left[(\alpha\sigma^6(1-\lambda)+r^6)^{\frac{1}{6}}\right]
\end{eqnarray}
where $V(r)$ denotes the ordinary electrostatic or Lennard-Jones potential in the GROMOS force field in the A-state ($\lambda=0$) and in the B-state ($\lambda=1$) \cite{Gunsteren94}.
The parameter $\alpha$ has been chosen to 0.7 and the radius of interaction value $\sigma$ was 0.3.\\
The free energy landscape for the PEO chain has been calculated by the metadynamics method presented in Ref.~\cite{Laio02}. A history dependent biasing potential in form of Gaussian hills 
was applied to the 
molecule during the simulation which helps to overcome energetic barriers.
Details of the method can be found elsewhere \cite{Laio02,Laio08,SmiatekWHM}.
The simulations have been conducted by the program plug-in PLUMED \cite{Plumed}. The Gaussian hills were set each 2 ps with a height of 0.1 kJ/mol and a width of 0.25 nm.
\section{4. Results and Discussion}
\subsection{4.1 Chymotrypsin inhibitor II}
\begin{figure}[h!]
 \includegraphics[width=0.5\textwidth]{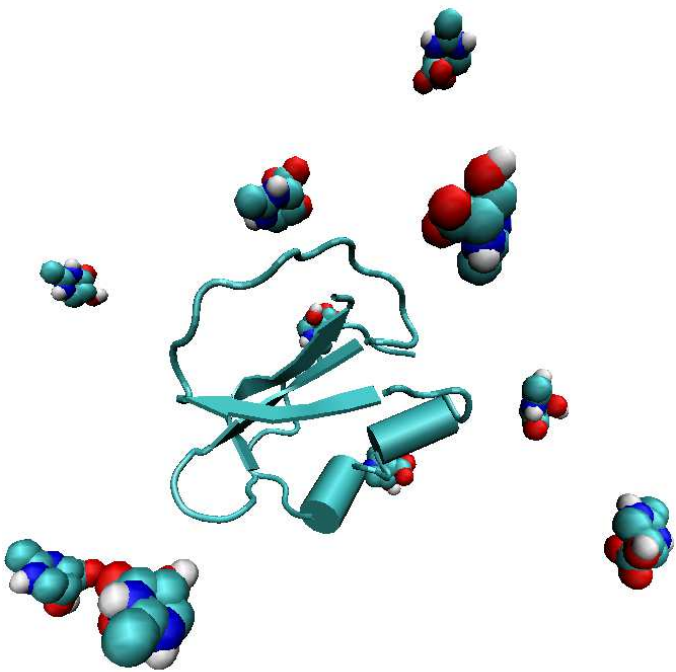}
\caption{Simulation snapshot of Chymotrypsin inhibitor II in presence of hydroxyectoine.
}
\label{fig2}
\end{figure}

We start this section by discussing the results for Chymotrypsin inhibitor II in presence of hydroxyectoine. 
A snapshot of a typical configuration for CI2 in a 0.087 molar solution of hydroxyectoine is shown in Fig.~\ref{fig2}.
At a first glance it can be seen that direct interactions of the co-solvent with the protein are nearly absent. 
For a more quantitative description, we have calculated the average number of intermolecular hydrogen bonds between solvent and protein $n_H^{solv}$ and between the residues of the protein $n_H^{CI2}$ for two 
different concentrations of hydroxyectoine shown in 
Tab.~\ref{tab1}.\\ 
It is clearly visible that the values for $n_H$ are nearly identical for both concentrations. This becomes in particular true by regarding the values for the intermolecular hydrogen bond density, which is defined
by $\rho_{_{H}}^{solv} = <n_{H}^{solv}/\Sigma>$ with the total solvent accessible surface area $\Sigma$. The values are $\rho_{_H}^{solv}=1.93$ nm$^{-2}$ in presence and $\rho_{_H}^{solv}=1.94$ nm$^{-2}$ 
in absence of hydroxyectoine.\\ 
Furthermore only slight variations of the intramolecular hydrogen bonds $n_{H}^{CI2}$ can be observed. It can be concluded that
the results do not agree with the preferential exclusion \cite{Timasheff02} due to identical values for the hydrogen bond densities as well as with the standard transfer free energy model \cite{Rose2008} 
due to the negligible differences in the number of intramolecular hydrogen bonds.\\
However, the stabilization effect of hydroxyectoine is clearly evident by analyzing the average end-to-end radii $r_e$ whose values are also shown in Tab.~\ref{tab1}. 
It can be seen that the protein is in a more compact conformation in presence of hydroxyectoine which 
is in agreement to the results published in Refs.~\cite{Yu04,Yu07}.
Finally we have calculated the solvent accessible
hydrophilic $\Sigma^{phil}$ and hydrophobic surface area $\Sigma^{phob}$. It comes out that the hydrophilic surface area decreases while the hydrophobic area increases in a hydroxyectoine/water mixture.
These at a first glance astonishing results can be directly related to a change in the solvent properties and will be discussed in the last section.\\
\begin{table}
\caption{Average number of intermolecular hydrogen bonds $n_H^{solv}$ between the solvent and CI2 and intramolecular hydrogen bonds $n_H^{CI2}$ between the residues of CI2, the average end-to-end radius $r_e$ and the solvent accessible
surface hydrophilic/hydrophobic area $\Sigma^{phil/phob}$ for two concentrations $c$ of hydroxyectoine.}
\label{tab1}       
\begin{tabular}{lccccc}
\hline\noalign{\smallskip}
$c$ [mol/L] & $n_H^{solv}$ & $n_H^{CI2}$ & $r_e$ & $\Sigma^{phil}$ & $\Sigma^{phob}$\\
\noalign{\smallskip}\hline\noalign{\smallskip}
0.087 & $117.84\pm 0.01$ & $30.88\pm 0.05$ & $1.45 \pm 0.01$ & $37.7$ & $22.8$\\
\noalign{\smallskip}\hline\noalign{\smallskip} 
0 & $115.58\pm 0.05$ & $33.48\pm 0.04$ & $1.54 \pm 0.01$ & $38.1$ & $22.0$ \\
\noalign{\smallskip}\hline
\end{tabular}
\end{table}
In addition we have studied the number of hydrogen bonds between CI2 and hydroxyectoine. 
We have found an average value of $1.09 \pm 0.01$ hydrogen bonds between both solutes. 
This result is in agreement to a recent publication, which has reported the occurrence of direct binding between hydroxyectoine and lipid headgroups for high molar concentrations \cite{Galla11}. 
Nevertheless, regarding the large number of hydroxyectoine molecules, it can be concluded that preferential binding of hydroxyectoine 
which has been found as important \cite{Tolan2011} 
for zwitterionic co-solvents is a minor effect for protein stabilization.  
This becomes obvious by regarding the correlation coefficient between the end-to-end radius and the number of direct bonds with hydroxyectoine 
which is given by a negative value of $-0.012$. Hence it can be concluded that the stabilization of a compact state in presence of hydroxyectoine is mainly caused by indirect solvent interactions. 
\subsection{4.2 Polyethylene oxide}
\begin{figure}
 \includegraphics[width=0.5\textwidth]{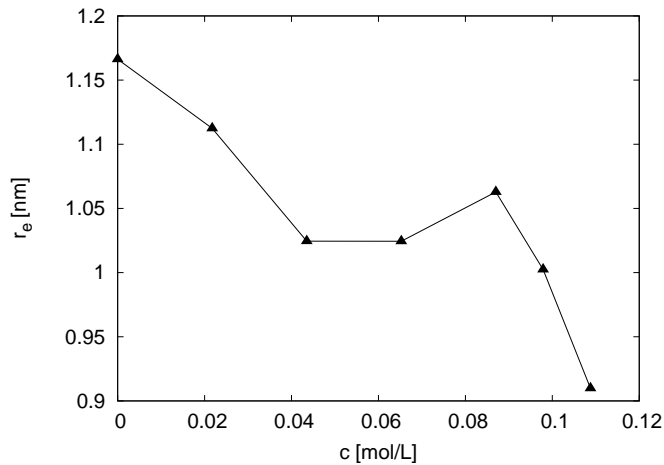}
\caption{End-to-end radius $r_e$ for the PEO chain in presence of varying hydroxyectoine concentrations $c$.
}
\label{fig3}
\end{figure}

To study the stabilization process in more detail, we have simulated a PEO chain in aqueous solution for various hydroxyectoine concentrations. PEO consists of polar and apolar atoms 
($\Sigma^{phil}/\Sigma^{phob} \approx 0.1$)
such that the stabilizing effects of co-solvents can be investigated by a chemically simple PEO molecule instead of complex proteins.\\  
We start this investigation by calculating the end-to-end radius $r_e$ for PEO in absence of hydroxyectoine and for a concentration of 0.11 mol/L.
The average end-to-end radius is $r_e=1.17\pm 0.01 $ nm in absence of hydroxyectoine and $r_e=0.91\pm 0.01$ nm for a 0.11 molar concentration.
Thus PEO shows a significant shrinkage of about 22\% for the end-to-end distance. 
To investigate the influence of direct interactions between PEO, hydroxyectoine and water molecules, 
we have calculated the number of intermolecular hydrogen bonds with water. It is nearly constant for all hydroxyectoine concentrations with a value of $n_{_H}^{solv} = 5.38 \pm 0.02$. 
The same is true for hydrogen bonds 
between hydroxyectoine and PEO with $n_{_H}^{PEO}=0.07 \pm 0.02$.  
Thus it is obvious that direct interactions between PEO and hydroxyectoine are absent.\\ 
The influence of varying hydroxyectoine concentrations on the end-to-end radius can be observed in Fig.~\ref{fig3}.
It becomes obvious that similar to CI2 the PEO chain is more compact
in presence of hydroxyectoine. This effect is drastically pronounced above a concentration of 0.09 mol/L.\\
\begin{figure}[h!]
 \includegraphics[width=0.5\textwidth]{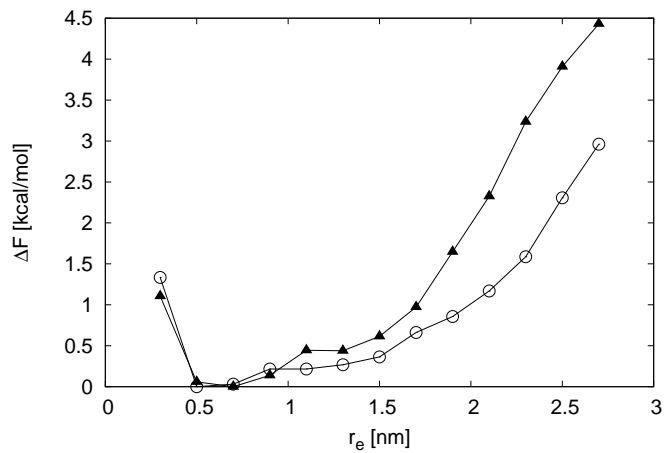}
\caption{Free energy landscape for the end-to-end radius of PEO for a 0.11 molar hydroxyectoine solution (triangles) and and in absence of  hydroxyectoine (circles).
}
\label{fig4}
\end{figure}
To study the swelling process energetically, we have calculated the free energy landscape for the end-to-end radius via metadynamics simulations.
A nearly identical method has been applied in a recent publication \cite{Sironi11}.
The landscape in presence (c=0.11 mol/L) and in absence of hydroxyectoine is presented in Fig.~\ref{fig4}.
An increase of the free energy can be observed in presence of compatible solutes
for larger values of the end-to-end radius. For a pure solution, $r_e$ values between 0.55 to 1.65 nm with an energy difference of $\Delta F \approx 0.3$ kcal/mol are energetically comparable.
In presence of hydroxyectoine, identical differences can be observed between $r_e=0.55$ and 0.85 nm. 
Values larger than 1 nm are energetically unfavorable in presence of the extremolyte. 
Combined with the above presented results, it can be concluded that the compact structure is stabilized due to an increase of the free energy barriers in presence of hydroxyectoine.
\subsection{4.3 Solvent properties}
\begin{figure}
 \includegraphics[width=0.5\textwidth]{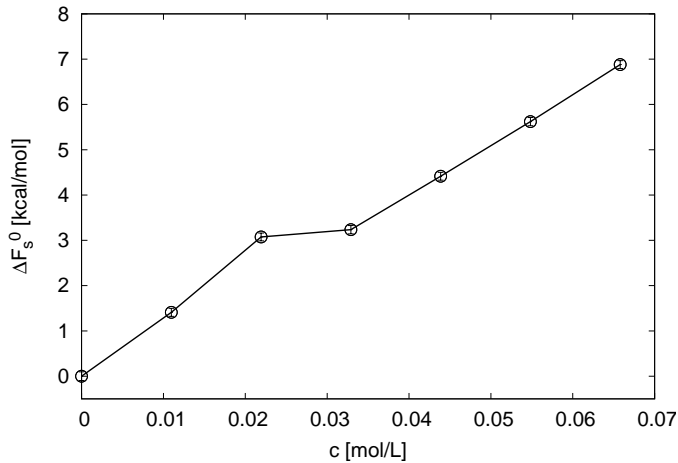}
\caption{Free solvation energy differences of ectoine for various concentrations of hydroxyectoine compared to the pure solution. 
}
\label{fig5}
\end{figure}

\begin{figure}
\includegraphics[width=0.5\textwidth]{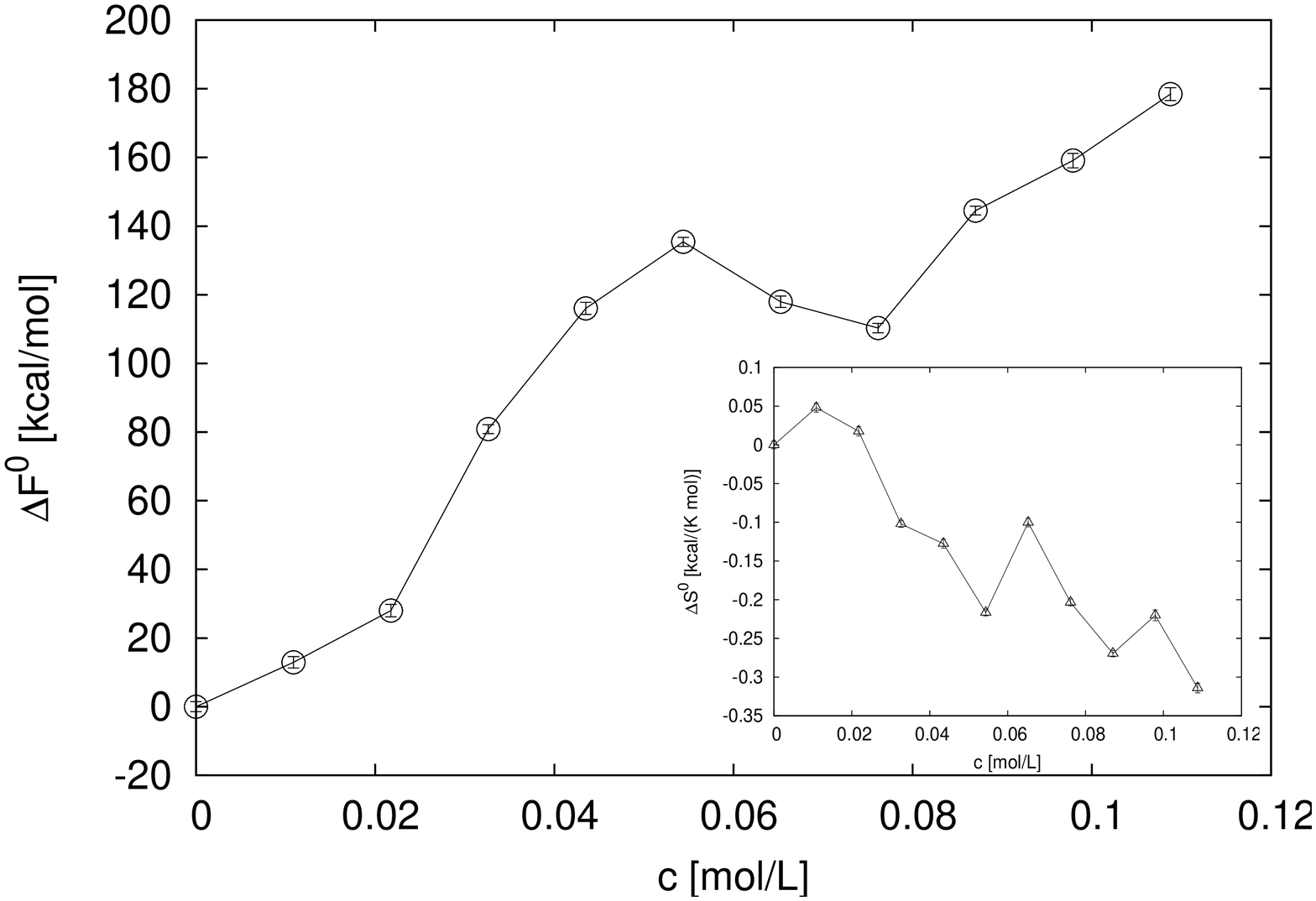}
 \caption{Excess free energy difference to the pure solution $\Delta F^0$ for various concentrations of hydroxyectoine. {\bf Inset:} Corresponding excess entropies $\Delta S^0$. 
}
\label{fig6}
\end{figure}

To study this effect in more detail, we have calculated the free solvation energy of ectoine, which is a zwitterionic molecule for various concentrations of hydroxyectoine. 
The results for the free solvation energy differences to a pure solution $\Delta F_{{s}}^0$ are shown in Fig.~\ref{fig5}.
It is worth to notice that the free solvation energy of ectoine is largely exothermic with a value of $\Delta F_s = -89.97 \pm 0.1$ kcal/mol. Thus it can be concluded that the solubility of ectoine
is decreased for larger concentrations of the co-solvent.
The linear dependence is obvious.  
Direct interactions between the co-solvent and ectoine are absent. Hence it becomes obvious that the main mechanism is indirectly mediated by the solvent such that the free solvation energy for polar residues or atoms
is significantly increased in presence of hydroxyectoine.\\
To study the consequences for the solvent in detail, we have performed simulations with varying concentrations of hydroxyectoine in absence of macromolecules or other solutes.\\
The excess free energy yields an estimate for the deviation of a real solution to the ideal state. It can be computed by the following equation
\begin{equation}
F^{ex} = F - F_{id} = -k_BT \log<e^{-V/k_BT}>
\end{equation}
where $V$ denotes the potential energy of the mixture and $k_BT$ the thermal energy \cite{Frenkel96}.
The excess free energy difference of a hydroxyectoine/water mixture to a pure solution of water is given by $\Delta F^0 = F^{ex}-F^{ex,0}$ where $F^{ex,0}$ is the excess free energy of the pure water. 
The results are shown in Fig.~\ref{fig6}. The increase of the excess free energy difference for larger hydroxyectoine concentrations is remarkable.\\ 
Furthermore the excess entropy difference is given by $\Delta S^0 = (\Delta U^0 -\Delta F^0)/T$ with $\Delta U^0 = U^{ex}-U^{ex,0}$ where $U$ is the total energy of the system. 
In agreement to the excess free energy differences, decreased entropies can be observed for larger concentrations of hydroxyectoine.
It can be concluded that due to the decreased entropy differences the solvent is locally more ordered in presence 
of hydroxyectoine in agreement to recent results \cite{Smiatek11}.\\
\begin{figure}
 \includegraphics[width=0.5\textwidth]{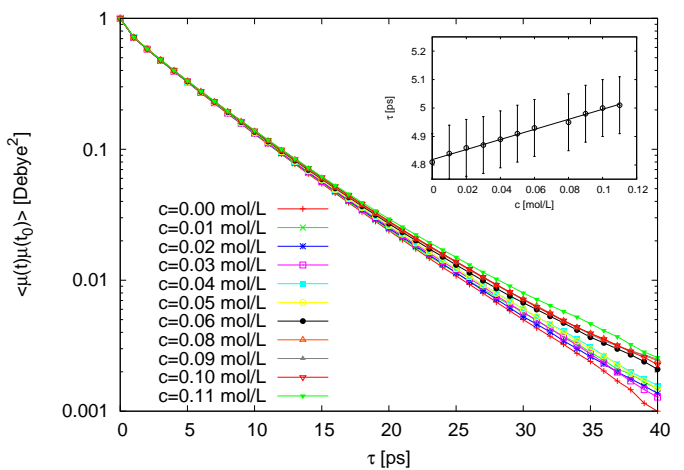}
\caption{Normalized dipole autocorrelation function for water molecules in varying hydroxyectoine concentrations. {\bf Inset:} Relaxation times $\tau$ for the different 
hydroxyectoine concentrations. The values can be fitted by regression to $\tau_0 = 4.82\pm 0.01$ ps with a slope of $1.77 \pm 0.05$ ps(mol/L)$^{-1}$.
}
\label{fig7}
\end{figure}
The effects of hydroxyectoine on the solvent properties can be also observed by the investigation of the dipolar autocorrelation function \cite{Spoel98} 
$<\vec{\mu}(t)\vec{\mu}(t_0)>\sim\exp(t/\tau)$ with the correlation time $\tau$  
averaged over all water molecules. The results are presented in Fig.~\ref{fig7}.
Although the effects are small, the linear dependence of the concentration on the relaxation times as well as the averaging effect over all molecules emphasizes the relative influence of 
hydroxyectoine. 
It is obvious that the presence of a high molar concentration of hydroxyectoine 
is directly related to increased relaxation times. Hence it can be concluded that the structural reorganization of water is diminished in presence of hydroxyectoine which validates the observed 
excess entropy values shown in Fig.~\ref{fig6}. 
Similar effects can be also observed in the context of hydrophobic hydration \cite{Engberts93} and the influence of disaccharides on water dynamics \cite{Verde11}.\\ 
\begin{figure}
 \includegraphics[width=0.5\textwidth]{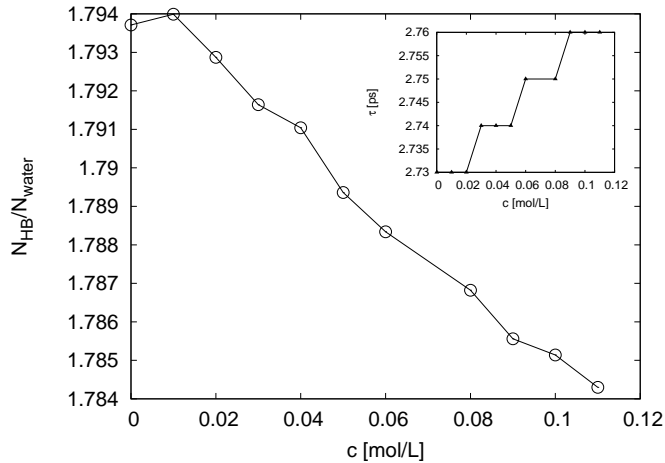}
\caption{Average number of hydrogen bonds of a single water molecule with other water molecules for varying hydroxyectoine concentrations. {\bf Inset:} Corresponding
lifetimes for hydrogen bonds. 
}
\label{fig8}
\end{figure}
The decrease in entropy is also supported by the results shown in Fig.~\ref{fig8} where the average number of hydrogen bonds for a single water molecule is presented. 
Although the variations are small, 
it can be assumed 
that the generalized net effect of a smaller number of hydrogen bonds for all water molecules compared to the zero hydroxyectoine concentration is significant. 
The above discussed increased relaxation times are additionally included in the hydrogen bond lifetimes which show an increase of 0.03 ps.\\
All the above presented results are related to a significant variation of the entropic contributions and the local ordering of the solution resulting in an increase of the free energy.\\   
\begin{figure}
 \includegraphics[width=0.5\textwidth]{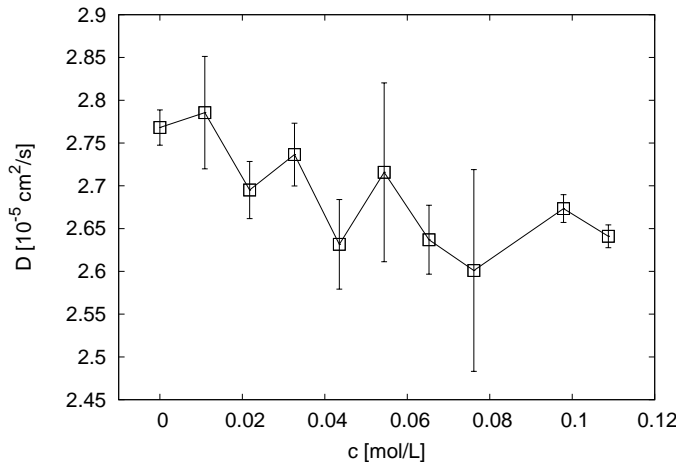}
\caption{Diffusion coefficient of water molecules for different hydroxyectoine concentrations. 
}
\label{fig9}
\end{figure}
A restriction in molecular motion becomes also obvious in Fig.~\ref{fig9} where the average diffusion constant $D$ for a water molecule in presence of varying hydroxyectoine concentrations is shown. 
It is evident that $D$ is smaller in presence of larger hydroxyectoine 
concentrations. It has to be noticed that this result is in agreement to Ref.~\cite{Yu04}, 
where the slowdown of water molecules for higher ectoine concentrations has been additionally observed. 
Thus we assume that the presence of hydroxyectoine results in a decreased water diffusivity in agreement to Ref.~\cite{Yu04}. 
Thus we conclude that hydroxyectoine changes the properties of the solvent such that unfolding of proteins becomes less favorable.
\section{5. Summary and Conclusion}
We have investigated the influence of compatible solutes on the structural properties of CI2 and PEO.
It has been found that in presence of hydroxyectoine the protein as well as the polymer show a significant preservation of the compact structure.
Energetic stabilization of compact states due to an increase of the free energy barriers has been observed by analyzing the results of metadynamics simulations. These findings are 
supported by an increase of the free solvation energy in presence of hydroxyectoine for polar ectoine molecules.\\ 
Furthermore direct stabilizing interactions like intra- and intermolecular hydrogen bonds are nearly absent 
as our results have shown. By studying the properties of a water-hydroxyectoine mixture we have indicated a significant increase of the excess free energy in combination with decreased excess entropy values. 
The observed entropy deficit can be validated by an increase of relaxational orientation times, increase of water hydrogen bond lifetimes as well as a decreased diffusivity.
The free energy increase can be also related to a lower average number of hydrogen bonds between water molecules.\\
By a combination of these observations, we propose the following mechanism which is responsible for the preservation of a proteins compact configuration. 
Direct interactions between hydroxyectoine and the macromolecule can be neglected due to the absence of intermolecular hydrogen bonds. 
Due to the increased values for the polar molecules free solvation energy, it can be assumed that a swelling of hydrophilic surface areas is less energetically favorable in presence of a highly concentrated 
solution of hydroxyectoine compared to a pure water solution.
This becomes also obvious by the values for the hydrophilic/hydrophobic solvent accessible surface area for CI2 shown in Tab.~\ref{tab1}. It is remarkable that the size of the hydrophobic area increases in presence 
of hydroxyectoine whereas the hydrophilic area decreases.\\ 
As a direct consequence, apolar hydrophobic molecules have decreased energy values for large hydroxyectoine concentrations as it has been validated for HO-CH$_2$-OH (data not shown).\\
This effect is mainly responsible for the stabilization of the proteins compact state.
It is widely accepted that protein folding to the compact state is initiated by a hydrophobic collapse \cite{Dobson98,Engberts93} where
hydrophobic residues are shielded in the interior whereas polar residues stay in direct contact with the solvent \cite{Dobson98,Engberts93}.\\ 
Considering the reverse direction, unfolding is correlated with a swelling of the total solvent accessible surface area. 
In the presence of hydroxyectoine, an increase of the hydrophilic surface area is energetically unfavorable. Compared to a pure solution, 
the compact native state is stabilized due to an increase of the free energy barriers such that swelling is suppressed.\\
It has to be noticed that the proposed mechanism is universal for all proteins with well-defined hydrophilic surface areas regardless of the chemical details.\\
The deficits for the free solvation energy in presence of compatible solutes are in good agreement to experimental results \cite{Roesgen05} and to the conclusions of the transfer free energy model \cite{Rose2008}
which assumes a decrease of solvent quality for large co-solvent concentrations. 
Hence the interaction of the macromolecules with the solvent becomes less favorable such that a shrinkage process due to solvophobic effects \cite{Rose2008} can be observed.
Detailed deviations 
from the transfer free energy model in our simulation are given by a negligible excess number of intramolecular hydrogen bonds to stabilize the proteins compact structure.\\ 
Our results validate that solvophobic effects are strong enough to stabilize the preservation of the compact native state as it can be also seen by the free energy landscapes shown in  Fig.~\ref{fig4}
and as it has been proposed in Ref.~\cite{Roesgen05}. 
Unfolding is energetically less preferable in presence of hydroxyectoine such that the equilibrium constant even at harsh
conditions is shifted towards the native structure. These results are in good agreement to experimental findings \cite{Roesgen05}.\\ 
Finally we are also able to explain the results of a recent publication \cite{Yu07}, where it has been found that the melting temperature for Met-Enkephalin remains unchanged in presence of ectoine. 
By detailed analysis of Met-Enkephalin, it comes out that the the ratio of the hydrophobic surface area to the total solvent accessible surface area is larger compared to CI2. 
Due to the discussion above, we conclude that the stabilizing effects of the co-solvent on the hydrophilic regions are less important for Met-Enkephalin. In addition it has been also validated that
a well-defined unfolding process for Met-Enkephalin is hard to observe due to low energy barriers between different conformations \cite{SmiatekWHM}.\\
Summarizing all results, we propose a simple mechanism in close agreement to the transfer free energy model to explain the stabilizing effect of extremolytes on the native state of proteins. 
Our results imply that the solvophobic effect is the main driving force to shift the equilibrium constant to the 
native form in absence of 
direct intra- and intermolecular interactions. The strategy of extremolyte production in microbacteriae under environmental stress can be 
explained by the proposed mechanism which is validated by significant variations of the solvent properties. 
\section{Acknowledgments}
The authors thank Oliver Rubner and Rakesh Kumar Harishchandra for enlightening discussions.
Financial support by the Deutsche Forschungsgemeinschaft (DFG) through the transregional collaborative research center
TRR 61 and the SFB 858 is gratefully acknowledged.

\end{document}